\documentclass{article}
\usepackage{amssymb}
\usepackage{amsmath}

\input{tcilatex}
\begin{document}

\title{Some Generalities About Generality}
\author{John D. Barrow \\
DAMTP, University of Cambridge\\
Wilberforce Rd., Cambridge CB3 0WA\\
UK}
\date{}
\maketitle

\begin{abstract}
We survey a variety of cosmological problems where the issue of generality
has arisen. This is aimed at providing a wider context for many claims and
deductions made when philosophers of science choose cosmological problems
for investigation. We show how simple counting arguments can be used to
characterise parts of the general solution of Einstein's equations when
various matter fields are present and with different spatial topologies.
Applications are described to the problem of singularities, static
cosmological models, cosmic no hair theorems, the late-time isotropisation
of cosmological models, and the number of parameters needed to describe a
general astronomical universe.
\end{abstract}

\section{Introduction}

The equations of general relativity and its extensions are mathematically
complicated and their general coordinate covariance offers special
challenges to anyone seeking exact solutions or conducting numerical
simulations. They are non-linear in a self-interacting (non-Abelian) way
because the mediator of the gravitational interaction (the graviton) also
feels the gravitational force. By contrast in an Abelian theory, like
electromagnetism, the photon does not possess the electric charge that it
mediates. As a result of this formidable complexity and non-linearity, the
known exact solutions of general relativity have always possessed special
properties. High symmetry, or some other simplifying mathematical property,
is required if Einstein's equations are to be solved exactly. General
solutions are out of reach.

This 'generality' problem has been a recurrent one in relativistic cosmology
from the outset in 1916 when Einstein \cite{einstat} first proposed a static
spatially homogeneous and isotropic cosmological model with non-Euclidean
spatial geometry in which gravitationally attractive matter is
counter-balanced by a positive cosmological constant. This solution turned
out to be unstable \cite{static}. Subsequently, the appearance of an
apparent 'beginning' and 'end' to simple expanding universe solutions led to
a long debate over whether these features were also unstable artefacts of
high symmetry or special choices of matter in the known cosmological
solutions, as Einstein thought possible. The quest to decide this issue
culminated in a new definition of such 'singularities' which allowed precise
theorems to be proved without the use of special symmetry assumptions. In
fact, by using the geodesic equations, their proofs made no use of the
Einstein equations \cite{HE}. Special solutions of Einstein's equations,
like the famous G\"{o}del metric \cite{godel} with its closed timelike
curves, also provoked a series of technical studies of whether its
time-travelling paths are a general feature of solutions to Einstein's
equations, or just isolated unstable examples. In the period 1967-1980 there
was considerable interest in determining whether the observed isotropy of
the microwave background radiation could be explained because it appeared to
be an unstable property of expanding universes \cite{mis1}. The mechanism of
'inflation', first proposed in 1981 by Guth \cite{guth}, provided a scenario
in which this conclusion could be reversed, and isotropy could be a stable
(or asymptotically stable) property of expanding universe solutions, by
widening the allowed conditions on the allowed forms of matter that could
dominate the expansion dynamics of the very early universe \cite{nohair},%
\cite{star}. Just to show how knowledge, fashion, and belief change, the
requirements on the density, $\rho $, and pressure, $p$, of matter content
needed for inflation to occur ($\rho +3p<0$) are exactly the opposite of
those assumed ($\rho +3p>0$) in the principal singularity theorems of
Penrose and Hawking \cite{HE, ear} in order to establish sufficient
conditions for a singularity (at least one incomplete geodesic) to have
occurred in our past.

In the study of differential equations, an exact solution is called \textit{%
stable} if small perturbations remain bounded as time increases; it is
called \textit{asymptotically stable} if the perturbations die away to zero
with increasing time. Our solar system is dynamically stable but not
asymptotically stable. Another useful pair of definitions are those
introduced by Hawking \cite{SWHgen} in 1971, who uses the same word in a
technically different way. He defines a 'stable' or 'open' property of a
dynamical system to be one that occurs from an open set (rather than merely
a single point) in initial data space. However, it is possible for a
property of a cosmological model to be stable but be of no physical
interest: it is a necessary but not a sufficient property for physical
relevance because the property in question could be stable only in open
neighbourhoods of initial data space describing universes with other highly
unrealistic properties (contraction or extreme anisotropy, for example). A
'generic' or 'open dense' property will be one that occurs near almost every
initial data set (that is, it is open dense on the space of all initial
data). A sufficient condition for a stable property to be of physical
interest is that it is generic in this sense \cite{BT}.

In this chapter we will discuss approaches to the problem of assessing
generality and some of the results that arise in typical and topical
cosmological problems. We will try to avoid significant technicalities.
There is a deliberate emphasis upon fundamental questions of interest to
philosophers of science rather than upon the astrophysical complexities of
the best-fit cosmological models or the galaxy of inflationary universe
models. Attention will be focussed on classical general relativity; aspects
of quantum cosmology will be treated in other chapters.

\section{General relativistic and Newtonian cosmology}

General relativity is a much larger theory than Newtonian gravity. It has
ten symmetric metric potentials, $g_{ab}$, instead of one Newtonian
gravitational potential,$\Phi $, and ten field equations (Einstein's
equations) instead of a single one (Poisson's equation to determine them
from material content of space and time. Newtonian gravity has a fixed time
and a fixed space geometry which is usually taken to be a monotonous linear
time plus a 3-d Euclidean space (although another fixed curved space could
be used simply by using the appropriate $\nabla ^{2}$ operator in Poisson's
equation).

Despite appearances, Newton's theory is not really complete and Newtonian
cosmology is not a well-posed theory \cite{gotz, norton}. Unlike general
relativity, it contains no propagation equations for the shear distortion
and the formulation of anisotropic cosmological models requires these to be
put in by hand. As a result the Newtonian description of an isotropic and
homogeneous cosmology looks exactly like general relativity \cite{mccrea}
because these shear degrees of freedom are necessary absent. This feature
manifests itself in results for the general asymptotic behaviour of the
Newtonian n-body problem in the unbound (expanding) case. Rigorous results
can be obtained for the moment of inertia (or radius of gyration), or
rotation of the total finite mass of n-bodies, but not for its shape \cite%
{saari},\cite{gotz}.

All solutions of Einstein's equations describe entire universes. The
relative sizes of the two theories means that infinitely many of these
general relativistic solutions possess no Newtonian counterpart. However,
there are also Newtonian 'universes' which have no counterpart in general
relativity. For example there are shear-free Newtonian solutions with
expansion and rotation: these cannot exist in general relativity \cite%
{ellis1}. More striking, there exist solutions of the Newtonian n-body
problem (for $n>4$) in which a system of point particles expands to infinite
size in finite time, undergoing an infinite number of oscillations in the
process \cite{xia}. For example, two counter-rotating binary pairs, all of
equal mass, with a lighter particle oscillating between their centres along
a line perpendicular to their orbital planes can expand to infinite size as
a result of an infinite number of recoils in a finite time! This is only
possible because Newtonian point particles can get arbitrarily close to one
another and so the $1/r^{2}$ forces between them an become arbitrarily
large. In general relativity this cannot happen. When two point particles of
mass $M$ approach closer than $4GM/c^{2}$ an event horizon forms around
them. This is a simple example of a form of 'cosmic censorship' that saves
us from the occurrence of an actual observable infinity, in Aristotle's
sense \cite{jbinf}, locally. In general relativity there is evidence that
under broad conditions there is a maximum force, equal to $c^{4}/G$, \cite%
{maxF} as well as the more fundamental a maximum velocity for information
transfer, $c$ : neither of these relativistic limits of velocity and force
strength exist in Newtonian theory.

\section{Generality -- some historic cases}

There have been a succession of cosmological problems where particular
solutions were found with striking properties that required further analysis
to determine whether those properties were general features of cosmological
solutions to the Einstein equations.

\subsection{Static universes}

The first isotropic and homogeneous cosmological model found by Einstein 
\cite{einstat} was a static universe with zero-pressure matter, positive
cosmological constant and a positive curvature of space. Subsequently, this
solution was shown to be unstable when it was perturbed within the family of
possible isotropic and homogeneous solutions of Einstein's equations by
Eddington and implicitly by Lema\^{\i}tre \cite{lem} who found the general
solutions of which Einstein's universe was a particular, and clearly
unstable, case. These demonstrations led to the immediate abandonment of the
static universe and, in Einstein's case, of the cosmological constant as
well \cite{blun}. It turns out that this stability problem is more
complicated than it appears and has only been completely explored, when
other forms of matter are present, quite recently. The static universe is
only unstable against small inhomogeneous perturbations on scales exceeding
the Jeans length when $p/\rho <1/5$. When $1\geq p/\rho >1/5,$ the Jeans
length for the inhomogeneities exceeds the size of the universe and so the
instability does not become Jeans unstable and amplify in time \cite{static}.

\subsection{Singularities}

There is a long and interesting history of attempts to interpret and avoid
'singularities' in the cosmological solutions of Einstein's equations. In
the first expanding solutions with zero-pressure matter found by Friedmann 
\cite{fried}, it appeared that there was a necessary beginning to the
expansion with infinite density at a finite time in the past and there could
(in spatially closed cases) also be an apparent end to the universe at a
finite time in the future. One response to these infinities, particularly by
Einstein, was to question whether they would remain if the family of
solutions was widened. First, Einstein asked whether the addition of
pressure would resist the compression to infinite density. Lema\^{\i}tre
showed that adding pressure actually made the problem worse by hastening the
appearance of the infinite density \cite{lem1}. The reason is a relativistic
one. Whereas in Newtonian physics any pressure resists gravitational
compression, in relativity pressure also gravitates because it is a form of
energy (and so has an equivalent mass via '$E=mc^{2}$') and increases the
compression (see also \cite{chandra}). Next, Einstein wondered whether it
was the perfect isotropy of the expanding universe solutions that was
responsible. If anisotropy was allowed then perhaps the compression would be
defocussed and the singularity avoided. Again, Lema\^{\i}tre was easily able
to show that simple anisotropic universes have the same types of singularity
and they are approached quicker than in the isotropic case \cite{lem1}.

These investigations by Lema\^{\i}tre amounted to tests of the stability of
the singularity occurrence within different wider sets of initial data. Many
cosmologists were convinced by these examples that singularities were
ubiquitous in these types of cosmology unless new forms of matter could be
found which resisted compression to infinite density. One such material was
the C-field of the steady state theory, introduced by Hoyle to describe
'continuous creation' of matter in a de Sitter universe that avoided the
high-density singularities of the Friedmann-Lema\^{\i}tre models. However,
all its null geodesics are past incomplete and so it is technically singular 
\cite{BTAP} (a feature that has been recently rediscovered in another
context \cite{borde}).

\ Later, in the period 1957-1966, a different approach was pursued for a
while by members of Landau's school in Moscow, notably by Khalatnikov and
Lifshitz \cite{KL}. Initially, they set out to show that singularities did
not occur in the general solution of the Einstein equations. Their argument,
which was incorrect, was that because singularities arose in solutions that
were not general (like the Friedmann solutions or the anisotropic Kasner
universes) they would appear in the general solution. The circumstantial
evidence for this conclusion was the belief that the singularity arising in
these cosmological solutions was just a singularity in the coordinate system
used to describe the dynamics and so was unphysical (as is the 'singularity'
that arises at the North Pole of the Earth where the meridians intersect in
standard mapping coordinates). When it occurred you could change to a new
set of coordinates until they too became singular (as they always did) and
so on ad infinitum. Unfortunately, it is important to investigate what
happens in the limit of this process: a true physical singularity remains as
became increasingly clear when the problem was subjected to a different sort
of analysis. The singularity theorems of Hawking and Penrose \cite{HE} were
able to define sufficient conditions for the formation of a singularity by
adopting a definition of a singularity as an inextendible path of a particle
or light-ray in spacetime. These theorems made no reference to special
symmetries or the subtleties of coordinate choices. Singularities were where
time ran out: part of the edge of spacetime \cite{JB}. It remained to be
shown that these endpoints were caused by infinities in physical quantities.
To some extent this can be done but the full story is by no means complete,
even now. These theorems ended the argument about whether cosmological
singularities were physically real and general and later work by Belinskii,
Khalatnikov and Lifshitz refocussed upon finding the general behaviour near
a physical singularity \cite{BKL}.

It is important to stress that the singularities are \textit{theorems}, not 
\textit{theories}. They give \textit{sufficient} conditions for
singularities so if their assumptions are not all met this does not mean
that there is no inevitable singularity, merely that no conclusions can be
drawn. The interesting historical aspect which we signalled in the
introduction is that the sufficient conditions generally included the
requirement that the matter content of the universe obeys $\rho +3p>0$. We
no longer believe this inequality holds for all matter sources. Indeed, the
observations that the universe is accelerating could be claimed to show that
the assumption that $\rho +3p>0$ is false.

\subsection{Isotropisation}

After the discovery in 1967 of the high level of isotropy in the CMB
temperature distribution \cite{part} there was a long effort to explain why
this was the case. Up until then cosmologists had assumed an isotropic and
homogeneous background universe and regarded the presence of small
inhomogeneities (like galaxies) as the major mystery requiring a simple
explanation. The discovery of the CMB isotropy placed created a new
perspective in which it was the high isotropy and uniformity of the assumed
background universe that was the major mystery. A new approach, proposed by
Misner and dubbed 'chaotic cosmology' sought to show that general
cosmological initial conditions would end up leading to an isotropically
expanding universe after more than about 10 billion years \cite{mis1}. This
programme had an interesting methodological aspect. If it could be shown
that almost all initial conditions (subject to some weak conditions of
physical reasonableness) would lead to isotropic universe then observations
of the isotropy level could not tell us anything about the initial
conditions: memory of them would have been erased by the expansion.

In studying whether this idea could work it was again the issue of
generality that was crucial. It was asking whether isotropically expanding
universes were stable or even asymptotically stable attractors at late
times. Two types of analysis were performed. The first just asked whether
anisotropic cosmologies would approach isotropy at late times if they just
contain zero-pressure matter and radiation. \ The second, which Misner
proposed, was to ask what happened if dissipative stresses could arise
because of the presence of collisionless particles, like neutrinos and
gravitons, at particular epochs in the very early universe. Perhaps large
initial anisotropies could be damped out by these dissipative processes,
leaving an isotropically expanding universe?

Several interesting approaches to these questions were developed. On the
physical level Barrow and Matzner \cite{BM} showed that the chaotic
cosmology philosophy could not work in general because the dissipation of
anisotropies and inhomogeneities must produce heat radiation, in accord with
the second law of thermodynamics. The earlier dissipation occurred the
larger entropy per baryon produced and the observed entropy per baryon today
(of about $10^{9}$) placed a prohibitively strong bound on how much
anisotropy could have been damped out during the history of the universe.
The presence of particle horizons with proper radii proportional to $t$ in
the early universe also placed a major constraint on the damping of any
large scale inhomogeneities by causal processes. Misner \cite{mix} attempted
to circumvent this by discovering the remarkable possibility that spatially
homogeneous universes of Bianchi type IX (dubbed the 'Mixmaster' universe
because of this property) could potentially allow light to travel around the
universe arbitrarily often on approach to $t=0$ as a result of their chaotic
dynamics. Unfortunately, this horizon removal mechanism was ineffective in
practice because of the improbability of the horizonless dynamical
configurations and the fact that only about 20 chaotic oscillations of the
scale factor could have occurred between the Planck time ($10^{-43}$s) and
the present \cite{DNov}. There was also the concern that if general
relativistic cosmology was a well-behaved initial-value problem then one
could always concoct anomalously anisotropic universe today that could be
evolved back to their initial conditions at any arbitrary early time \cite%
{CS}. These would provide counter examples to the chaotic cosmology scheme
although one has to be careful with this argument because the counter
examples could all have physically impossible initial conditions, and in
fact they often do \cite{JB2}.

In 1973 Collins and Hawking \cite{CH} carried out some interesting stability
analyses of isotropic universes to discover if isotropy was a stable
property of homogeneous initial data. The results found were widely
discussed and reported at a semi-popular level but needed to be treated
cautiously because of the fine detail in the theorems. They reported that
for ever-expanding universes isotropic expansion was 'unstable' but if
attention was narrowed to spatially flat initial data with zero-pressure
matter then isotropy was 'stable'. The cosmological constant was assumed to
be zero. The definition of stability used was in fact \textit{asymptotic
stability} and so the proof that isotropy was unstable just meant that
anisotropies did not tend to zero as $t\rightarrow \infty $. In fact, closer
analysis showed that in general $\sigma /H\rightarrow $ constant in open
universes (and it is impossible for $\sigma /H$ to grow asymptotically) with 
$\rho +3p>0$. This means that isotropy is \textit{stable, }although not%
\textit{\ asymptotically stable,} \cite{late,JB3}.\textit{\ }The other
technicality, is that this result is a consequence the fact that these open
universe become vacuum (or spatial curvature dominated) at late times. The
behaviour of the anisotropy it therefore an asymptotic property of vacuum
cosmologies and doesn't tell us anything about the past history of the
universe at redshifts $z>z_{c}$, where $z_{c}$ $<O(1)$ is the redshift where
the expansion becomes curvature dominated. These stability results therefore
did not help us understand the sort of initial data that could give rise to
high isotropy after about 10 billion yeas of expansion.

\subsection{Cosmic no-hair theorems}

Amid all this interest in explaining the isotropy of the universe there was
one prescient approach by Hoyle and Narlikar \cite{hoy} that predated the
discovery of the high isotropy of the CMB. In 1963 they pointed out that in
the standard big bang model the isotropy and average uniformity of the
universe was a mystery but in the steady state universe it would be
naturally explained. To support this claim they showed that the de Sitter
universe is stable against scalar, vector and tensor perturbations. Thus, a
steady state universe, described by the de Sitter metric of general
relativity would always display high isotropy and uniformity. In fact, had
they only known it they could have predicted that the only spectrum of
perturbations consistent with the steady state universe is the constant
curvature spectrum with constant (small) metric perturbations on all scales
that is observed with high accuracy today -- although ironically via
perturbations in the CMB whose existence the steady state model could not
explain). Any departure from this spectrum, with $\delta \rho /\rho \propto
L^{-2}$ on length scale $L$ would either create divergent metric potential
perturbations as $L\rightarrow \infty $ or $L\rightarrow 0$.

The idea of the inflationary universe provided a new type of explanation for
the observed isotropy and uniformity of the universe from general initial
conditions, but with one important difference from past expectations -- the
isotropy and homogeneity was predicted to be \textit{local}. The
inflationary universe theory proposed that there was a finite interval of
time, soon after the apparent beginning of the expansion (typically at $\sim
10^{-35}$s in the original conception of the theory), when the expansion of
the universe would accelerate due to the presence of very slowly evolving
scalar fields of the sort that appeared in new theories of high-energy
physics. These would contribute stresses with $\rho +3p<0$ and cause the
expansion scale factor to accelerate. When this occur the expansion rapidly
approaches isotropy. Anisotropies fall off very rapidly and isotropic
expansion is asymptotically stable. In the most likely scenario, where $%
p=-\rho $, the expansion behaves temporarily like Hoyle's steady state
universe and grows exponentially in time. However, the inflation needs to an
end, and this can happen if the scalar fields responsible decay into
ordinary particles and radiation with $\rho +3p>0$. When this happens the
usual decelerating expansion is resumed but with anisotropies so diminished
in amplitude that they remain imperceptibly small late late times \cite%
{nohair,star}.

Inflation works by taking a small patch of the universe that is small enough
for light signals to cross it at an early time $t_{I}$ and expanding it so
dramatically (exponentially in time) that it grows larger than the entire
visible universe today in the short period during which inflation occurs.
Thus the isotropy and high uniformity of the visible universe today is a
reflection of the fact that it is the expanded image of a region that was
small enough to be coordinated by light-like transport processes and damping
when inflation occurred. If inflation had not occurred, and the expansion
had merely continued along its standard decelerating trajectory then the
initially smooth and isotropic region would not have expanded significantly
by the present time, $t_{0}$. Here is a simple calculation of how this
happens.

Suppose the preset temperature of the CMB is $T_{0}=3K$ and when inflation
occurred is was $T_{I}=3\times 10^{28}K.$ Then, since $T\propto a^{-1},$ the
scale factor has increased by a factor of $T_{I}/T_{0}=10^{28}$. At time $%
t_{I}$, the horizon size is equal to $d(t_{I})=2ct_{I}$ where $t_{I}\simeq
10^{-35}s,$ so a horizon-sized region of size $2ct_{I\text{ }}$at $t_{I}$
would only have expanded to a size $2\times 3\times 10^{10}$cms$^{-1}\times
10^{-35}s\times 10^{28}=6\times 10^{3}$cm by the present day. This is not of
any relevance for explaining isotropy and uniformity over scales of order $%
ct_{0}\sim 10^{28}$cm today. However, suppose inflation occurs at $t_{I}$
and inflates the expansion scale factor by a factor of $e^{N}$. We will now
be able to enlarge the causally connected region of size $2ct_{I}$ up to a
scale of $e^{N}\times 6\times 10^{3}$cm$.$This will exceed the size of the
visible universe today if $e^{N}\times 6\times 10^{3}\gtrsim 10^{28}$. This
is easily possible with $N>60$. If the expansion is exponential with $%
a(t)\propto e^{Ht}$ then we only need inflation to last from about $%
10^{-35}s $ until $10^{-33}s$ in order to effect this. The regularity of the
universe is therefore explained without any dissipation taking place. A very
tiny smooth patch is simply expanded to such an extent that its smooth and
isotropic character is reflected on the scale of the entire universe today.
It is very likely (just as it is more likely that a randomly chosen positive
integer will be a very large one) that the amount of inflation that occurred
will be much large than 60 e-folds. Yet, the result is to predict that the
universe will be uniform on the average out to the inflated scale $%
e^{N}\times 6\times 10^{3}$ but may be rather non-uniform if we could see
further. In some variants of the theory many other fundamental features of
the universe (values of constants of Nature, space dimensions. laws of
physics) are different beyond the inflated scale as well. While there have
always been overly positivistic philosophers who have cautioned against
simply assuming that the unobserved part of the (possibly infinite) universe
is the same on average as the observed part, this is the first time there
has been a positive prediction that we should not expect them to be the same.

The result of a sufficiently long period of accelerated expansion is to
drive the local expansion dynamics of the universe is to drive the expansion
dynamics towards the isotropic de Sitter expansion, with asymptotic form of
the metric of the form ($\alpha ,\beta =1,2,3$) \cite{star}

\begin{eqnarray*}
ds^{2} &=&dt^{2}-g_{\alpha \beta }dx^{\alpha }dx^{\beta }, \\
g_{\alpha \beta } &=&\exp [2Ht]a_{\alpha \beta }(\mathbf{x})+b_{\alpha \beta
}(\mathbf{x})+\exp [-Ht]c_{\alpha \beta }(\mathbf{x})+...,
\end{eqnarray*}%
where $H$ is the constant Hubble rate, with $3H^{2}=\Lambda ,$ and $%
a_{\alpha \beta }(\mathbf{x}),b_{\alpha \beta }(\mathbf{x})$ and $c_{\alpha
\beta }(\mathbf{x})$ are arbitrary symmetric spatial functions. The Einstein
equations allow only two of the $a_{\alpha \beta }$ and two of the $%
c_{\alpha \beta }$ to be freely specifiable and all the $b_{\alpha \beta }$
are determined by them. Thus there are four independently arbitrary spatial
functions specifying the solution on a spacelike surface of constant in
vacuum. Notice the spatial functions $a_{\alpha \beta }(\mathbf{x})$ at
leading order in the metric ($H$ is a constant though). This is why the
metric only approaches de Sitter locally, exponentially rapidly inside the
event horizon of a geodesically moving observer. If radiation is black body (%
$p=\rho /3$) is added then a further 4 arbitrary spatial functions are
required (three for the normalised 4-velocity components and one for the
density) and

\begin{eqnarray*}
\rho &\propto &\exp [-4Ht], \\
u_{0} &\rightarrow &1,u_{\alpha }\propto \exp [Ht]c_{\alpha ;\beta }^{\beta
}, \\
c_{\alpha }^{\alpha } &=&0.
\end{eqnarray*}%
Hence, we see that the 3-velocity $V^{2}=u_{\alpha }u^{\alpha }$ tends to a
constant as $t\rightarrow \infty $. The asymptotic state is therefore de
Sitter plus a constant (or 'tilted') velocity field which affects the metric
at third order (via $c_{\alpha \beta }(\mathbf{x})$).This is easy to
understand physically if we consider a large rotating eddy that expands with
the universe and has angular velocity $\omega =Va^{-1}$. Its angular
momentum is $Ma^{2}\omega \propto (\rho a^{3})a^{2}(Va^{-1})$ and this is
conserved as the universe expands. since the radiation density falls as $%
\rho \propto a^{-4},$ we have $V$ constant as $a\rightarrow \infty $.

The number of free spatial functions specifying this asymptotic solution is
8 in the case with radiation (and the same holds when any other perfect
fluid matter is present). In the next section we will show that this is
characteristic of a part of the general solution of Einstein's equations.

\ In conclusion we see that a finite period of accelerated expansion is able
to drive the expansion towards isotropy from a very large class of initial
conditions (not all initial conditions, since the universe must not, for
example recollapse before a period of accelerated expansion begins). We have
just discussed the most extreme form of accelerated expansion with constant
Hubble expansion rate, $H$, and $a\propto \exp [Ht]$ but similar conclusions
hold for power-law inflation, with $a\propto t^{n},n>1,$ and intermediate
inflation, with $a\propto \exp [At^{n}],$ with $A>0,$and $0<n<1$ constants.
The key conceptual point is that explains the present isotropy without
dissipating initial anisotropies in the way that the chaotic cosmology
programme imagined and so it evades the Barrow-Matzner entropy per baryon
constraint \cite{BM}. Instead it drives the initial inhomogeneities far
beyond the visible horizon today and the stress driving the acceleration
dominates over all forms of anisotropy at large expansion volumes and times.
The earlier analyses of the stability of isotropic expansion by Collins and
Hawking, and others, \cite{CH}, had restricted attention to forms of matter
in the universe with $\rho +3p>0$ and always assumed $\Lambda =0$ because
there was no reason to think otherwise at that time. As a result, they had
excluded the possibility of accelerated expansion which can solve the
isotropy problem without any dissipation occurring if it can arise for a
finite period of time in the early universe .

\subsection{The initial value problem}

The attempts to explain the isotropy of the universe from arbitrary initial
conditions gave rise to another interesting perspective that is worth
highlighting. General relativity is an initial value problem an so for
'well-behaved' cosmological solutions this means that the present state of
the universe described by any solution of Einstein's equations is a
continuous function of some 'initial data' at any past time. In a technical
sense it might appear that given any state of the universe today -- highly
anisotropic, for example -- then there exists some initial data set that
evolves to give that state \ regardless of the action of any damping
effects. Hence, there could never be a theory that could explain the actual
state of the universe today as the result of evolution from any (or almost
any) initial conditions. The problem with this argument is that the initial
conditions that do evolve to counter-factual cosmological states at late
times may arise only from initial data states that are completely unphysical
in some respect \cite{JB2}. Take a simple example of a Bianchi type I
anisotropic universe. The anisotropy energy density and radiation energy
density fall as

\begin{eqnarray*}
\sigma ^{2} &=&\sigma _{0}^{2}(1+z)^{6}, \\
\rho _{\gamma } &=&\rho _{\gamma 0}(1+z)^{4}.
\end{eqnarray*}%
We can choose values of the constant $\sigma _{0}^{2}$ so that the
universe's expansion is dominated by anisotropy today -- just pick $\sigma
_{0}^{2}=\rho _{\gamma 0}$ $\simeq 10^{-34}$gmcm$^{-3}$ to specify the
initial data. However, if we run this apparent counter-example back to the
time when the radiation temperature is $T_{pl}\simeq 10^{32}K$ $=3(1+z_{pl})$
when its energy density equals the Planck density, $10^{94}$gmcm$^{-3}$, we
require the anisotropy energy density to be $10^{64}$ times larger than the
Planck energy density at that time -- a completely unphysical situation.
Alternatively, if we had taken the anisotropy energy density to be the
Planck density at $z_{pl}$ then we have the strange initial condition that
the radiation density is $10^{64}$ times smaller despite all forms of energy
being in quantum gravitational interaction at that time.

This is a (deliberately) dramatic example but the basic problem with the
argument is one that one can find with other arguments regarding the
generality of more complicated outcomes in cosmology. For example, there
have been claims (and claims to the contrary) that inflation is not generic
for Friedmann universes containing scalar fields with a quadratic
self-interaction potential \cite{meas}. The claim is based on using the
Hamiltonian measure in the phase space for the dynamics to show that the
bulk of the initial data measure is for solutions which don't inflate. This
type of initial data corresponds to solutions with huge initial kinetic term
($\dot{\phi}^{2}$) which dominate the potential $V(\phi )=m^{2}\phi ^{2}$ by
a huge factor so that the potential never comes to dominate the dynamics by
any pre-specified epoch. However, this doesn't look very natural because it
requires the two forms of energy density to differ by an enormous factor
when one of them equals the Planck energy density (above which we know
nothing about what happens since general relativity, quantum mechanics and
statistical mechanics all break down). The better course is to not let any
energy density exceed the Planck value but (surprisingly) this appears to be
controversial.

\section{Naive function counting}

There have been several attempts to reduce the description of the
astronomical universe to the determination of a small number of measurable
parameters. Typically, these will be the free parameters of a well defined
cosmological model that uses the smallest number of constants that can
provide a best fit to the available observational evidence. Specific
examples are the popular characterisations of cosmology as a search for
'nine numbers' \cite{rrob}, 'six numbers' \cite{rees}, or the six-parameter
minimal $\Lambda CDM$ model used to fit the WMAP \cite{wmap} and Planck data
sets \cite{Planck}. In all these, and other, cases of simple parameter
counting there are usually many simplifying assumptions that amount to
ignoring other parameters or setting them to zero; for example, by assuming
a flat Friedmann background universe or a power-law variation of density
inhomogeneity in order to reduce the parameter count and any associated
degeneracies. The assumption of a power-law spectrum for inhomogeneities
will reduce a spatial function to two constants, while the assumption that
the universe is described by a Friedmann metric plus small inhomogeneous
perturbations both reduces the number of metric unknowns and converts
functions into constant parameters. In this paper we are going provide some
context for the common minimal parameter counts cited above by determining
the total number of spatial functions that are needed to prescribe the
structure of the universe if it is assumed to contain a finite number of
simple matter fields. We are not counting fundamental constants of physics,
like the Newtonian gravitation constant, the coupling constants defining
quadratic lagrangian extensions of general relativistic gravity, or the 19
free parameters that define the behaviours of the 61 elementary particles in
the standard 3-generation $U(1)\times SU(2)\times SU(3)$ model of particle
physics. However, there is some ambiguity in the status in some quantities.
For example, as to whether the dark energy is equivalent to a true
cosmological constant (a fundamental constant), or to some effective fluid
or scalar field, or some other emergent effect \cite{shaw} Some fundamental
physics parameters, like neutrino masses, particle lifetimes, or axion
phases, can also play a part in determining cosmological densities but that
is a secondary use of the cosmological observable. Here we will take an
elementary approach that counts the number of arbitrary functions needed to
specify the general solution of the Einstein equations (and its
generalisations).This will give a minimalist characterisation that can be
augmented by adding any number of additional fields in a straightforward
way. We will also consider the count in higher-order gravity theories as
well as for general relativistic cosmologies. We enumerate the situation in
spatially homogeneous universes in detail so as to highlight the significant
impact of their spatial topology on evaluations of their relative generality.

Let us move on to a more formal discussion of how to specify the generality
of solutions to Einstein's equations by counting the number of free
functions (or constants) that a given solution or approximate solutions
contains. In view of the constraint equations and coordinate covariances of
the theory this requires a careful accounting.

The cosmological problem can be formulated in general relativity using a
metric in a general synchronous reference system \cite{LL}. Assume that
there are $F$ matter fields which are non-interacting and each behaves as a
perfect fluid with some equation of state $p_{i}(\rho _{i})$, $i=1,...F$.
They will each have a normalised 4-velocity field, $(u_{a})_{i}$, $%
a=0,1,2,3. $ These will in general be different and non-comoving. Thus each
matter field is defined on a spacelike surface of constant time by $4$
arbitrary functions of three spatial variables, $x^{\alpha }$ since the $%
u_{0}$ components are determined by the normalisations $(u_{a}u^{a})_{i}=1$.
This means that the initial data for the $F$ non-interacting fluids are
specified by $4F$ functions of three spatial variables. If we were in an $N$%
-dimensional space then each fluid would require $N+1$ functions of $N$
spatial variables and $F$ fluids would require $(N+1)F$ such functions to
describe them in general.

The 3-d metric requires the specification of $6$ $g_{\alpha \beta \text{ \ }%
} $and $6$ $\dot{g}_{\alpha \beta }$ for the symmetric spatial $3\times 3$
metric in the synchronous system but these may be reduced by using the $4$
coordinate covariances of the theory and a further $4$ can be eliminated by
using the $4$ constraint equations of general relativity. This leaves $4$
independently arbitrary functions of three spatial variables \cite{LL} which
is just twice the number of degrees of freedom of the gravitational spin-2
field. The general transformation between synchronous coordinate systems
maintains this number of functions \cite{LL}. This is the number required to
specify the general vacuum solution of the Einstein equations in a $3$%
-dimensional space. In an $N$-dimensional space we would require $N(N+1)$
functions of $N$ spatial variables to specify the initial data for $%
g_{\alpha \beta \text{ \ }}$and $\dot{g}_{\alpha \beta }$. This could be
reduced by $N+1$ coordinate covariances and $N+1$ constraints to leave $%
(N-2)(N+1)$ independent arbitrary functions of $N$ variables \cite{JDB}.
This even number is equal to twice the number of degrees of freedom of the
gravitational spin-2 field in $N+1$ dimensional spacetime .

When we combine these counts we see that the general solution in the
synchronous system for a general relativistic cosmological model containing $%
F$ fluids requires the specification of $(N-2)(N+1)+F(N+1)=(N+1)(N+F-2)$
independent functions of $N$ spatial variables. If there are also $S$
non-interacting scalar fields, $\phi _{j}$, $j=1,..,S,$ present with self
interaction potentials $V(\phi _{j})$ then two further spatial functions are
required ($\phi _{j}$ and $\dot{\phi}_{j}$) to specify each scalar field and
the total becomes $(N+1)(N+F-2)+2S$. For the observationally relevant case
of $N=3$, this reduces to $4(F+1)+2S$ spatial functions.

For example, if we assume a simple realistic scenario in which the universe
contains separate baryonic, cold dark matter, photon, neutrino and dark
energy fluids, all with separate non-comoving velocity fields, but no scalar
fields, then $F=$ $5$ and our cosmology needs $24$ spatial functions in the
general case. If the dark energy is not a fluid, but a cosmological constant
with constant density and $u_{i}=\delta _{i}^{0},$ then the dark energy
'fluid' description reduces to the specification of a single constant, $\rho
_{DE}=\Lambda /8\pi G$, rather than $4$ functions and reduces the total to $%
21$ independent spatial functions. However, if the cosmological constant is
an evolving scalar field then we would have $F=4$ and $S=1$, and now $22$
spatial functions are required. Examples of full function asymptotic
solutions were found for perturbations around de Sitter space-time by
Starobinsky \cite{star}, the approach to 'sudden' finite-time singularities 
\cite{sudd} by Barrow, Cotsakis and Tsokaros \cite{BC}, and near
quasi-isotropic singularities with $p>\rho $ 'fluids' by Heinzle and Sandin 
\cite{heinz}.

These function counts of $21$-$24$ should be regarded as lower bounds. They
do not include the possibility of a cosmological magnetic field or some
other unknown matter fields. They also treat all light ($<<1MeV$) neutrinos
as if they are identical (heavy neutrinos can be regarded as CDM if they
provide the largest contribution to the matter density but if they are not
responsible for the dominant dark matter then they should be counted as a
further contribution to $F$). If there are matter fields which are not
simple fluids with $p(\rho )$ -- for example an imperfect fluid possessing a
bulk viscosity or a gas of free particles with anisotropic pressures -- then
additional parameters are required to specify them. There can still be
overall constraints -- a trace-free energy-momentum tensor, for example, in
the cases of electric and magnetic fields or Yang-Mills fields -- and we
would just count the number of independent terms in the total
energy-momentum tensor \cite{JB4}.

In the case of the Planck or WMAP mission data analyses, $6$ constants are
chosen to define the standard (minimal) $\Lambda CDM$ model. For WMAP \cite%
{wmap}, these are the present-day Hubble expansion rate, $H_{0}$, the
densities of baryons and cold dark matter, the optical depth, $\tau $, at a
fixed redshift, and the amplitude and slope of an assumed power-law spectrum
of curvature inhomogeneities on a specified reference length scale. This is
equivalent to including three matter fields (radiation, baryons, cold dark
matter) but the standard $\Lambda CDM$ assumes zero spatial curvature, $k$, 
\textit{ab initio} so a relaxation of this would add a curvature term or a
dark energy field, because when $k\neq 0$ the latter could no longer be
deduced from the other densities and the critical density (defined by $H_{0}$%
). The light neutrino densities are assumed to be calculable from the
radiation density using the standard cosmological thermal history, so there
are effectively $F=5$ matter fields (with $k$ set to zero in the base model)
and a metric time derivative determined by $H$). All deviations from
isotropy and homogeneity enter only at the level of perturbation theory and
are characterised by the spectral amplitude and slope on large scales; the
amplitude on small scales ('acoustic peaks' in the power spectrum) is
determined from that on large scales by an $e^{-2\tau }$ damping factor
determined by the optical depth parameter $\tau $. The Planck mission
parameter choice is equivalent to this \cite{Planck}.

Although a general solution of the Einstein equations requires the full
complement of arbitrary functions, different parts of the general solution
space can have behaviours of quite different complexity. For example, when $%
N\leq 9$ there are homogeneous vacuum universes which are dynamically
chaotic \cite{BKL, JBcha} but the chaotic behaviour disappears when $N\geq
10 $\ even though the number of arbitrary constants remains maximal for each 
$N$ \cite{chaos}. Hence, the dynamical complexity can fail to be captured by
the function-counting approach.

\subsection{Einstein's 'strength'}

As an interesting historical aside, we should mention Einstein's attempt to
study the power of mathematical formalism to describe physical theories by
ascribing to them a numerical measure of their predictive power, which he
called the 'strength' of a system of differential equations. It was to be
measured by the number of free pieces of initial data needed to determine
the general solutions of the equations. Einstein believed that 'The smaller
the number of free data consistent with the system of field equations, the
'stronger' is the system. It is clear that in the absence of any other
viewpoint from which to select the equations, one will prefer a 'stronger'
system to a less strong one.' \cite{ein} This was the method Einstein \
proposed to follow in his quest for a unified field theory (how different to
the methodology that led to all his past great successes). The enumeration
of the strength of a system of equations for $d$ variables began by
expanding an analytic function of these variables in a Taylor about a point
and noting that at $n^{th}$ order the total number of terms in the expansion
is

\begin{equation*}
\binom{n+d-1}{n}\equiv \frac{(n+d-1)!}{n!(d-1)!}.
\end{equation*}

If there are field equations which ensure that when the function is
specified arbitrarily on a $d-1$ dimensional (spatial) surface then those in
the remaining (temporal) dimension are determined by them, then only $\left( 
\begin{array}{c}
d+1 \\ 
n%
\end{array}%
\right) $ of the Taylor series coefficients remain arbitrary. The fraction
of coefficients that remain free is therefore

\begin{equation*}
\frac{\left( 
\begin{array}{c}
d-1 \\ 
n%
\end{array}%
\right) }{\left( 
\begin{array}{c}
d \\ 
n%
\end{array}%
\right) }=\frac{d-1}{n+d-1}.
\end{equation*}

In the case of Einstein's equations we have coordinate covariances and
constraint equations to use to reduce the count of free functions. The
resulting strength turns out to be identical to the count of independent
pieces of initial data for the metric and its first derivative that we have
just described, giving a strength of $4$ in vacuum. A similar count can be
done for Maxwell's equations (which have the same strength), or other
equations of mathematical physics. A fuller discussion is given in refs \cite%
{schutz, mari}.

\section{More general gravity theories}

There has been considerable interest in trying to explain the dark energy as
a feature of a higher-order gravitational theory that extends the lagrangian
of general relativity in a non-linear fashion \cite{BO, far, ferr, clif}.
This offers the possibility of introducing a lagrangian that is a function
of $L=f(R,R_{ab}R^{ab})$ of the scalar curvature $R$ and/or the Ricci scalar 
$R_{ab}R^{ab}$ in anisotropic models, with the property that it contributes
a slowly varying dark energy-like behaviour at late times without the need
to specify an explicit cosmological constant. However, these higher-order
lagrangian theories (excluding the Lovelock lagrangians in which the
variation of the higher-order terms contribute pure divergences \cite{Love}
and so the field equations are always 2nd order in any spatial dimension)
all have 4th-order field equations in 3-dimensional space when $f$ $\neq
A+BR,$ with $A,B$ constants. This means that the initial data set for such
theories is considerably enlarged because we must specify $\ddot{g}_{\alpha
\beta \text{ }}$ and $\dddot{g}_{\alpha \beta }$ in addition to $g_{\alpha
\beta \text{ \ }}$and $\dot{g}_{\alpha \beta }$. In $N$ space dimensions,
this results in a further $N(N+1)$ functions of $N$ variables and so a
general cosmological model with $F$ fluids and $S$ scalar fields requires a
specification of $2(N^{2}-1)+F(N+1)+2S=(N+1)(F+2N-2)$ $+2S$ independent
arbitrary spatial functions. For $N=3,$this is $16+4F+2S$. General
relativity with $4$ matter fields plus a cosmological constant requires $20$
spatial functions plus one constant, in general, whereas a higher-order
gravity theory with $4$ matter fields and no scalar fields (and no
cosmological constant because it should presumably emerge from the metric
behaviour) requires the specification of $32$ spatial functions.

\section{Reducing functions to constants}

The commonest simplification used to reduce the size of the cosmological
characterisation problem is to turn the spatial functions into constants.
This simplification will be an exact if the universe is assumed to be
spatially homogeneous. The set of possible spatially homogeneous and
isotropic universes with natural topology is based upon the classification
of homogeneous 3-spaces created by Bianchi \cite{Bi, taub,mac, wain}
(together with the exceptional case of Kantowski-Sachs-Kompanyeets-Chernov
with $S^{1}\times S^{2}$ topology \cite{KS},\cite{KC} which we will ignore
here as it displays non-generic behaviour).

The most general Bianchi type universes are those of types $%
VI_{h},VII_{h},VIII$ and $IX$. \ Of these, only types $VII_{h\text{ }}$and $%
IX$, respectively, contain open and closed isotropic Friedmann subcases.
These most general Bianchi types are all defined by $4$ arbitrary constants
in vacuum plus a further $4$ for each non-interacting perfect fluid source.
Therefore, in 3-dimensional spaces, the most general spatially homogeneous
universes containing $F$ fluids are defined by $4(1+F)$ arbitrary \textit{%
constants}. This suggests that they might be the leading order term in a
linearisation of the general inhomogeneous solution in the homogeneous
limit. However, things might not be so simple. The 4-function space of
solutions to Einstein's models like type $IX$ with compact spaces has a
conical structure at points with Killing vectors and so linearisation about
the points must control an infinite number of spurious linearisations
(associated with all the tangents that can be drawn through the point of the
cone but don't run down the side of the cone) that are not the leading-order
terms in any convergent series expansion of a true solution \cite{mar, BT}.

The Bianchi classification of spatially homogeneous universes derives from
the classification of the group of isometries with 3-dimensional subgroups
that act simply transitively on the manifold. Intuitively, these give
cosmological histories that look the same to observers in different places
on the same hypersurface of constant time.

The Bianchi types are subdivided into two classes \cite{EM}: Class A
contains types\emph{\ }$%
I(1+F),II(2+3F),VI_{0}(3+4F),VII_{0}(3+4F),VIII(4+4F) $\emph{\ }and $%
IX(4+4F),$\ while Class B contains types $%
V(1+4F),IV(3+4F),III(3+4F),VI_{-1/9}(4+3F),VI_{h}(4+4F)$\ and $%
VII_{h}(4+4F). $ The brackets following each Roman numeral of the Bianchi
type geometry contain the number of constants defining the general solution
when $F$ non-interacting perfect fluids, each with $p>-\rho $, are present,
so $F=0$ defines the vacuum case. For example, Bianchi type I denoted by $%
I(1+F)$ is defined by one constant in vacuum (when it is the Kasner metric)
and one additional constant for the value of the density when each matter
field is added. For simplicity, we have ignored scalar fields here, but to
include them simply add $2S$ inside each pair of brackets. The Euclidean
metric geometry in the type $I$ case requires $R_{0\alpha }=0,$ identically,
and so the $3$ non-comoving velocities (and hence any possible vorticity)
must be identically zero. This contains the zero-curvature Friedmann model
as the isotropic (zero parameter) special case. In the next simplest case,
of type $V$, the general vacuum solution was found by Saunders \cite{sau}
and contains one parameter, but each additional perfect-fluid adds $4$
parameters because it requires specification of a density and three non-zero 
$u_{\alpha }$ components. The spatial geometry is a Lobachevsky space of
constant negative isotropic curvature. The isotropic subcases of type $V$
are the zero-parameter Milne universe in vacuum and the $F$-parameter open
Friedmann universe containing $F$ fluids.

In practice, one cannot find exact homogeneous general solutions containing
the maximal number of arbitrary constants because they are too complicated
mathematically, although the qualitative behaviours are fairly well
understood, and many explorations of the observational effects use the
simplest Bianchi I or V models (usually without including non-comoving
velocities) because they possess isotropic 3-curvature and add only a simple
fast-decaying anisotropy term (requiring one new constant parameter) to the
Friedmann equation. The most general anisotropic metrics which contain
isotropic special cases, of types $VII$ and $IX$, possess both expansion
anisotropy (shear) and anisotropic three-curvature. Their shear falls off
more slowly (logarithmically in time during the radiation era) and the
observational bounds on it are much weaker \cite{CH},\cite{dln},\cite{bbn},%
\cite{bjs},\cite{JB2,skew}.

\section{Some effects of the topology of the universe}

So far, we have assumed that the cosmological models in question have the
'natural' topology, that is $R^{3}$ for the 3-dimensional flat and
negatively curved spaces and $S^{3}$ for the closed spaces. However, compact
topologies can also be imposed upon flat and open universes to make their
spatial volumes finite and there has been considerable interest in this
possibility and its observational consequences for optical images of
galaxies and the CMB, \cite{ellis, top, top2}.

The classification of compact negatively-curved spaces is a challenging
mathematical problem. When compact spatial topologies are imposed on
spatially flat and open homogeneous cosmologies it produces a major change
in their relative generalities and the numbers of constants needed to
specify them in general.

The most notable consequences of a compact topology on 3-dimensional
homogeneous spaces is that the Bianchi universes of types $IV$ and $VI_{h}$
no longer exist at all and open universes of Bianchi types $V$ and $VII_{h}$
must be isotropic with spaces that are quotients of a space of constant
negative curvature, as required by Mostow's Rigidity theorem \cite%
{Ash,fagu,BK1,BK2,K}. The only universes with non-trivial structure that
differs from that of their universal covering spaces are those of Bianchi
types $I,II,III,VI_{0},VII_{0}$ and $VIII$. The numbers of parameters needed
to determine their general cosmological solutions when $F$ non-interacting
fluids are present and the spatial geometry is compact are now given by $%
I(10+F),II(6+3F),III(2+N_{m}^{\prime }+F),VI_{0}(4+4F),VII_{0}(8+4F)$ and $%
VIII(4+N_{m}+4F)$, again with $F=0$ giving the vacuum case, as before, and
an addition of $2S$ to each prescription if $S$ scalar fields are included.
Here, $N_{m}$ is the number of moduli degrees of freedom which measures of
the complexity of the allowed topology, with $N_{m}\equiv 6g+2k-6\equiv
N_{m}^{\prime }-2g$, where $g$ is the genus and $k$ is the number of conical
singularities of the underlying orbifold \cite{BK1,BK2}. It can be
arbitrarily large.

The rigidity restriction that compact types $V$ and $VII_{h}$ must be
isotropic means that compactness creates general parameter dependencies of $%
V(F)$ and $VII_{h}(F)$ which are the same as those for the open isotropic
Friedmann universe, or the Milne universe in vacuum when $F=0$.

The resulting classification is shown in Table 1, \cite{JB2014}. We see that
the introduction of compact topology for the simplest Bianchi type $I$
spaces produces a dramatic increase in relative generality. Indeed, they
become the most general vacuum models by the parameter-counting criterion.
An additional $9$ parameters are required to describe the compact type $I$
universe compared to the case with non-compact Euclidean $R^{3}$ topology.
The reason for this increase is that at any time the compact 3-torus
topology requires $3$ identification scales in orthogonal directions to
define the torus and $3$ angles to specify the directions of the vectors
generating this lattice plus all their time-derivatives. This gives $12$
parameters, of which $2$ can be removed using a time translation and the
single non-trivial Einstein constraint equation, leaving $10$ in vacuum
compared to the $1$ required in the non-compact Kasner vacuum case.

The following general points are worth noting:

(i) The imposition of a compact topology changes the relative generalities
of homogeneous cosmologies;

(ii) The compact flat universes are more general in the parameter-counting
sense than the open or closed ones;

(iii) Type $VIII$ universes, which do not contain Friedmann special cases
but can in principal become arbitrarily close to isotropy are the most
general compact universes.

The most general case that contains an isotropic special case is that of
type $VII_{0}$ -- recall that the $VII_{h}$ metrics are forced to be
isotropic so open Friedmann universes now become asymptotically stable \cite%
{BK1} and approach the Milne metric whereas in the non-compact case they are
merely stable and approach a family of anisotropic vacuum plane waves \cite%
{late}. This peculiar hierarchy of generality should be seen as a reflection
of how difficult it is to create compact homogeneous spaces supporting these
homogeneous groups of motions.

Table 1: \textit{The number of independent arbitrary constants required to
prescribe the general 3-dimensional spatially homogeneous Bianchi type
universes containing }$F$\textit{\ perfect fluid matter sources in cases
with non-compact and compact spatial topologies. The vacuum cases arise when 
}$F=0$\textit{. If }$S$ scalar \textit{fields are also present then each
parameter count increases by }$2S$\textit{.} \textit{The type }$IX$\textit{\
universe does not admit a non-compact geometry and compact universes of
Bianchi types }$IV$\textit{\ and }$VI_{h}$ \textit{do not exist. Types }$III$%
\textit{\ and }$VIII$\textit{\ have potentially unlimited topological
complexity and arbitrarily large numbers of defining constants parameters
through the unbounded topological parameters }$N_{m}\equiv 6g+2k-6$\textit{\
and }$N_{m}^{\prime }=N_{m}+2g$\textit{, where }$g$\textit{\ is the genus
and }$k$\textit{\ is the number of conical singularities of the underlying
orbifold \cite{JB2014}.}

\bigskip 
\begin{tabular}{|c|c|c|}
\hline
Cosmological & \multicolumn{2}{|c|}{No. of defining parameters with $F$
non-interacting fluids} \\ \cline{2-3}
Bianchi Type & Non-compact topology & Compact topology \\ \hline
$I$ & $1+F$ & $10+F$ \\ \hline
$II$ & $2+3F$ & $6+3F$ \\ \hline
$VI_{0}$ & $3+4F$ & $4+4F$ \\ \hline
$VII_{0}$ & $3+4F$ & $8+4F$ \\ \hline
$VIII$ & $4+4F$ & $4+N_{m}+4F$ \\ \hline
$IX$ & $-$ & $4+4F$ \\ \hline
$III$ & $3+4F$ & $2+N_{m}^{\prime }+F$ \\ \hline
$IV$ & $3+4F$ & $-$ \\ \hline
$V$ & $1+4F$ & $F$ \\ \hline
$VI_{h}$ & $4+4F$ & $-$ \\ \hline
$VII_{h}$ & $4+4F$ & $F$ \\ \hline
\end{tabular}

\section{Inhomogeneity}

The addition of inhomogeneity turns the constants defining the cosmological
problem into functions of three space variables. For example, we are
familiar with the linearised solutions for small density perturbations of a
Friedmann universe with natural topology which produces two functions of
space that control temporally growing and decaying modes. The function of
space in front of the growing mode is typically written as a power-law in
length scale (or wave number) and so has arbitrary amplitude and power index
(both usually assumed to be scale-independent constants to first or second
order) which can fitted to observations. Clearly there is no limit to the
number of parameters that could be introduced to characterise the density
inhomogeneity function by means of a series expansion around the homogeneous
model (and the same could be done for any vortical or gravitational-wave
perturbation modes) but the field equations would leave only 8 independent
functions. Further analysis of the function characterising the radiation
density is seen in the attempts to measure and calculate the deviation of
its statistics from gaussianity \cite{NG} and to reconstruct the past
light-cone structure of the universe \cite{M}. Any different choice of
specific spatial functions to characterise inhomogeneity in densities or
gravitational waves requires some theoretical motivation. What happens in
the inhomogeneous case if open or flat universes are given compact spatial
topologies is not known. As we have just seen, the effects of topology on
the spatially homogeneous anisotropic models was considerable whereas the
effects on the overall evolution of isotropic models (as opposed to the
effects on image optics) is insignificant. It is generally just assumed that
realistically inhomogeneous universes with non-positive curvature (or
curvature of varying sign) can be endowed with a compact topology and, if
so, this places no constraints on their dynamics. However, both assumptions
would be untrue for homogeneous universes and would necessarily fail for
inhomogeneous ones in the homogeneous limit. it remains to be determined
what topological constraints arise in the inhomogeneous cases. They could be
weaker because inhomogeneous anisotropies can be local (far small in scale
than the topological identifications) or they could be globally constrained
like homogeneous anisotropies. Newtonian intuitions can be dangerous because
compactification of a Newtonian Euclidean cosmological space seems simple
but if we integrate Poisson's equation over the compact spatial volume we
see that the total mass of matter must be zero. This follows from Poisson's
equation since

\begin{equation*}
0=\int_{V}\nabla ^{2}\Phi dV=4\pi G\int_{V}\rho dV=4\pi GM,
\end{equation*}%
where $V$ is the compact spatial volume, $M$ the total mass, and $\Phi $ is
the Newtonian gravitational potential.

\ In practice, there is a divide between the complexity of inhomogeneity in
the universe on small and large scales. On large scales there has been
effectively no processing of the primordial spectrum of inhomogeneity by
damping or non-linear evolution. Its description is well approximated by
replacing a smooth function by a power-law defined by 2 constants, as for
the microwave background temperature fluctuation spectrum or the 2-point
correlation function of galaxy clustering. Here, the defining functions may
be replaced by statistical distributions for specific features, like peak or
voids in the density distribution. On small scales, inhomogeneities that
entered the horizon during the radiation era can be damped out by photon
viscosity or diffusion and may leave distortions in the background radiation
spectrum as witness to their earlier existence. The baryon distribution may
provide baryon acoustic oscillations which yield potentially sensitive
information about the baryon density \cite{wmap, Planck}. On smaller scales
that enter the horizon later, where damping and non-linear self-interaction
has occurred, the resulting distributions of luminous and dark matter are
more complicated. However, they are correspondingly more difficult to
predict in detail and numerical simulations of ensembles of models are used
to make predictions down to the limit of reliable resolution. Predicting
their forms also requires a significant extension of the simple, purely
cosmological enumeration of free functions that we have discussed so far.
Detailed physical interactions, 3-d hydrodynamics, turbulence, shocks,
protogalaxy shapes, magnetic fields, and collision orientations, all
introduce additional factors that may increase the parameters on which
observable outcomes depend. The so called bias parameter, equal to the ratio
of luminous matter density to the total density, is in reality a spatial
function that is being used to follow the ratio of two densities because one
(the dark matter) is expected to be far more smoothly distributed than the
other. All these small scale factors combine to determine the output
distribution of the baryonic and non-baryonic density distributions and
their associated velocities.

\subsection{Links to Observables}

The free spatial functions (or constants) specifying inhomogeneous
(homogeneous) metrics have simple physical interpretations. In the most
general cases the $4$ vacuum parameters can be thought of as giving two
shear modes (ie time-derivatives of metric anisotropies) and two parts of
the anisotropic spatial curvature (composed of ratios and products of metric
functions). In the simplest vacuum models of type $I$ and $V$ the
three-curvature is isotropic an there is only one shear parameter. It
describes the allowed metric shear and in the type $V$ model a second
parameter is the isotropic three-curvature (which is zero in type $I$) --
just like $k$ in the Friedmann universe models. When matter is added there
is always a single $\rho ($or $p$) for each perfect fluid and up to three
non-comoving fluid velocity components. If the fluid is comoving, as in type 
$I$ only the density parameter is required for each fluid; in type $V$ there
can also be $3$ non-comoving velocities. The additional parameters control
the expansion shear anisotropy, anisotropic 3-curvature. They may all
contribute to temperature anisotropy in the CMB radiation but the observed
anisotropy is determined by an integral down the past null cone over the
shear (effectively the shear to Hubble rate ratio at last scattering of the
CMB), rather than the Weyl curvature modes driven by the curvature
anisotropy (which can be oscillatory \cite{wain2}, and so can be
periodically be very small even though the envelope is large), while the
velocities contribute dipole variations. Thus, it is difficult to extract
complete information about all the anisotropies from observations of the
lower multipoles of the CMB alone in the most general cases \cite{WS, M, Nil}%
.

At present, the observational focus is upon testing the simplest possible $%
\Lambda CDM$ model, defined by the smallest number (six) of parameters. As
observational sensitivity increases it will become possible to place
specific bounds or make determinations of the full spectrum of defining
functions (or constants), or at least to confirm that they remain
undetectably small as inflation would lead us to expect. In an inflationary
model they can be identified with the spatial functions defining the
asymptotic expansion around the de Sitter metric \cite{star}.

There have also been interesting studies of the observational information
needed to determine the structure of our past null cone rather than
constant-time hypersurfaces in the Universe \cite{ell}, extending earlier
investigations of the links between observables and general metric
expansions by McCrea \cite{mc} and by Kristian and Sachs \cite{ks}.

The high level of isotropy in the visible universe, possibly present as a
consequence of a period of inflation in the early universe \cite{guth}, or
special initial conditions \cite{spec, JB2, heinz, pen, swh}, is what allows
several of the defining functions of a generic cosmological model to be
ignored on the grounds that they are too small to be detected with current
technology. An inflationary theory of the chaotic or eternal variety, in
which inflation only ends locally, will lead to some complicated set of
defining functions that exhibit large smooth isotropic regions within a
complicated global structure which is beyond our visual horizon and
unobservable (although not necessarily falsifiable within a particular
cosmological model). However, despite the success of simple cosmological
theories in explaining almost all that we see in the universe, it is clear
that there is an under-determination problem: we cannot make enough
observations to specify the structure of space-time and its contents, even
on our past light cone, let alone beyond it. It is not a satisfactory
methodology to use observations to construct a description of space-time.
Rather, we proceed by creating parametrised descriptions that follow from
solutions of Einstein's equations (or some other theory) and then constrain
the free parameters by using he observational data. Despite the widespread
lip-service paid to Popper's doctrine of falsification as a scientific
methodology, its weaknesses are especially clear in cosmology. It assumes
that all observations and experimental results are correct and unbiased --
that what you see is what you get. In practice, they are not and you never
know whether observational data is falsifying a theory, or is based on wrong
measurements, or subject to some unsuspected selection effect \cite{JB5}.
All that observational science can ever do is change the likelihood of a
particular theory being true or false. Sometimes the likelihood can build up
(or down) to such an extent that we regard a theory being tested (like the
expansion of the universe) as 'true' or (like cold fusion) as 'false'.

\ Acknowledgements. Support from the STFC (UK) and the JTF Oxford-Cambridge
Philosophy of Cosmology programme grant is acknowledged.

\end{document}